*Original Article*

# Antibactrial Activity of *Asteriscus graveolens* Methanolic Extract: Synergistic Effect with Fungal Mediated Nanoparticles against Some Enteric Bacterial Human Pathogens

Ibtihal Nayel ALrawashdeh[a], Haitham Qaralleh[b*], Muhamad O. Al-limoun[a], Khaled M. Khleifat[a]

[a]Biology Department, Mutah University, Mutah, Karak, 61710, Jordan; [b]Department of Medical Laboratory Sciences, Mutah University, Mutah, Karak, 61710, Jordan*Corresponding Author: Haitham Qaralleh, [b]Department of Medical Laboratory Sciences, Mutah University, Mutah, Karak, 61710, Jordan

*Corresponding author: haitham@mutah.edu.jo

**Abstract:** Antibactrial activity of *Asteriscus graveolens* methanolic extract and its synergy effect with fungal mediated silver nanoparticles (AgNPs) against some enteric bacterial human pathogen was conducted. Silver nanoparticles were synthesized by the fungal strain namely *Tritirachium oryzae* W5H as reported early. In this study, MICs of AgNPs against *E. aerogenes*, *Salmonella* sp., *E. coli* and *C. albicans* in order were 2.13, 19.15, 0.08 and 6.38 µg/mL, respectively, while the MICs of *A. graveolens* ethanolic extract against the same bacteria were 4, 366, >3300 and 40 µg/mL, respectively. The MIC values at concentration less than 19.15 and 40 µg/ml indicating the potent bacteriostatic effect of AgNPs and *A. graveolens* ethanolic extract. Increasing in IFA was reported when Nitrofurantion and Trimethoprim were combined with Etoh extract with maximum increase in IFA by 6 and 12 folds for, respectively. Also, 10 folds increasing in IFA was reported when trimethoprim was combined with AgNPs: Etoh extract. But, there were no synergistic effect between the antifungal agents (Caspofungin and Micafungin) combined with AgNPs and or *A. graveolens* ethanolic extract against *C. albicans*. The potent synergistic effect of *A. graveolens* ethanolic extract and/or NPs with the conventional antibiotics is novel in inhibiting antibiotics resistant bacteria. In this study, remarkable increasing in the antibacterial activity, when the most resistant antibiotics combined with *A. graveolens* ethanolic extract and/or NPs was reported.

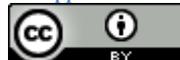

## INTRODUCTION

Nanotechnology rest on the synthesis and applications of nanoparticles in varied areas such as chemistry, physics, biology, and medicine (Pereira et al. 2014; Joanna et al. 2018). The use of microorganisms for synthesis of nanoparticles have emerged as a simple and viable alternative to more complex chemical synthetic procedures to obtain silver nanoparticles (AgNPs) (Siddiqi et al 2018). AgNPs have shown excellent bactericidal properties against a wide range of microorganisms (Naqvi et al. 2013). Nanoparticles can exhibit size-related properties significantly different from those of either fine particles or bulk materials (Duncan 2011; Durán et al., 2016). There has been worldwide development of multidrug-resistant bacteria because of above- or underuse of antibiotics (Rai et al. 2017; Khleifat,et al., 2019). These bacteria, as well as other types of microorganisms, have developed resistance to antibiotics, and consequently, this problem is now documented as an emergent global challenge (Roca et al. 2015). To manage the problem of multidrug resistance, new and different antimicrobials are required (Allahverdiyev et al. 2011). In this context, the use of silver nanoparticles in conjunction with different antibiotics and Etoh extract (Etoh) may exert antibacterial synergy, leading to the emergence of novel approaches for therapy. *Asteriscus graveolens* is an endemic Middle Eastern medicinal plant, located in extreme desert environments; its major Etoh extract components were identified as *cis*-chrysanthenyl acetate, myrtenyl acetate and kessane ( Tayeh et al. 2018).

Recent studies reported great antifungal and anticancer activities of its ethanol extract of *A. graveolens*. In addition, the ethanol extract of *A. graveolens* showed various degrees of antibacterial activity depending on the bacterial strain (Aouissi et al. 2018).

Nanoparticle–antibiotic combinations minimize the degree of combined agents in the dosage, which lowers noxiousness and raises antimicrobial potentials (Naqvi et al. 2013). Furthermore, due to this conjugation, the concentrations of antibiotics can be increased at the region of antibiotic–microbe contact and subsequently facilitate the binding between microorganisms and antibiotics (Allahverdiyev et al. 2011; Al-Asoufi et al., 2017).

The aim of the present work is to study the broad-spectrum antibacterial activity of AgNPs used singly and in combination with 6 different antibiotics or ethanol extract of *A. graveolens* leaves.

The effects of the bio-reduced silver nanoparticles when synergized with different antibiotics and the plant ethanol extract were investigated against three bacterial strains using the disc diffusion and microdilution methods. The combinations utilized in this study were (1), antibiotic plusAgNPs, (2), antibiotic plus ethanol extract of *A. graveolens* plants and (3), antibiotic plus AgNPs plus ethanol extract of plant. This is the first report using *Tritirachiumoryzae* as source of AgNPs and of the antibacterial capacity achieved by combining AgNPs and plant ethanol extractwith each tested antibiotic.

**MATERIALS AND METHODS**
**Materials, Growth media, reagents and equipment's**
The fungal strain used in this project was isolated from soil samples collected from olive oil mill at Al-Karak province, south of Jordan. The fungal strain was identified to the species level by ITS sequencing (GENWIZ, USA). Sequence similarity analysis with NCBI database was performed and then the sequence was registered in NCBI database and accession number was obtained.

**Plant material**
Aerial part of *A.graveolens* were collected in May and June from Dhana Natural Reserve (DNR), South of Jordan in 2018. The plants was identified as previously reported (Ahmed et al., 1991; Sacchetti et al., 2005).

**Bacterial strains**
Three bacterial species were used in this study including three Gram-negative bacteria: *E. coli* ATCC 11293, *E. aerogenes* ATCC 13048 *salmonella* ATCC 25664 and *one* fungal species *candida albican* ATCC 10231. The clinical isolates were obtained from ALkarak Hospital (Alkrak- Jordan). *E.coli* and *E.aerogenese* were obtained from patients with dermatitis (opportunistic skin infection) while *candida albican* was isolated from patient with high vaginal swab. All isolates were characterized by BIOMÉRIEUX VITEK® 2 SYSTEM. In addition, the other bacterial strains were all obtained as pure cultures from the Department of Biology, Faculty of Science, Mutah University, Mutah, Jordan.

**Preparation of silver nanoparticls**
For the biosynthesis of AgNPs, the approach presented by Jaidev and Narasimha, (2010)was taken into account. Furthermore, some changes such as the composition of liquid medium were set in place, as follows.The identified fungal isolate—*Tritirachium oryzaeW5H* was grown aerobically in a liquid media with the following chemical composition (w/v) 1.0% glucose, 1.0% yeast extract and 0.5% NaCl. Hundred-milliliter medium was inoculated with fungal spores at a rate of $2.0 \times 10^6$ and incubated at $33 \pm 4$ °C and speed of 150 rpm in an orbital shaker. After 72 h of incubation, the fungal biomass was filtered through Whatmann No.1 filter paper followed by extensive washing with deionized water. In a glass flask, 10 g (wet weight of biomass) was added to 100 ml of deionized water and incubated with shaking for 72 h (35°C at 150 rpm) (Khleifat et al., 2010). Then, the fungal biomass in the aqueous suspension was filtered through a Whatmann No.1 filter paper and the fungal filtrate was lastly obtained. Biosynthesis of AgNPs was attained via adding $AgNO_3$ to 100 ml of the fungal filtrate to achieve 1 mM concentration, which was incubated in the dark at 35°C and 150 rpm for 96-120 h. Control flask, without the AgNO3, was incubated at same conditions. Aliquots of the reaction solution were taken at 96-120 h of incubation for characterization of AgNPs by Ultraviolet-Visible (UV/VIS) spectrophotometer analysis.

**Ethanolic Extraction of *A. graveolens***
The fresh collected materials were air dried at room temperature in the shade. Then sample of 50 g of the dried plant material were extracted using 500ml ethanol and using shaker for 3 days. Then filter twice let it dry then we weigh

**Preparation of bacterial and fungal suspension**
Bacterial and fungal broth culture were made following the Laboratory Standards Institute (CLSI M7-A7, 2012). One bacterial and fungal colonies were cultivated into sterile 5 ml nutrient broth (NB) at 37°C for overnight period incubation. The resultant growth culture were adjusted to 0.5 McFarland Standard using sterile NB broth. The adjustment of bacterial and fungal suspensions to the density of the 0.5 McFarland standard was done spectrophotometrically at (A 620 nm) to obtain a final absorbance of 0.1 (Khleifat et al., 2014; Khleifat and Abboud, 2003; Khleifat et al., 2006a-c).

**Antimicrobial activity of *A.graveolens***
Stock solution of 100 mg/mL Etoh extract of *A. graveolens* in DMSO was prepared. For initial experiments of well diffusion method, 100 µl from the Etoh extract -DMSO solution was loaded to each well (10 mg per well). Meanwhile for disc diffusion method, 5µl of undiluted Etoh extract was applied to each disc (5µg/disc). In MIC assays, three-fold dilution series was performed from a stock solution of 100 mg/mL to obtain 3300, 1100, 370, 123, 40.7, 13.6, 4.53 and 1.5 µg/mL. The growth conditions were the same as previously mentioned. After 24 h, the zone of inhibition value of each sample was documented and compared to ethanolic extract alone to verify any synergistic effect among the tested antibiotics.

**Antimicrobial activity of AgNPs**
Stock solution of 1.726 mg/mL AgNPs was used. In well diffusion method, 100µl from the AgNPs solution was loaded to each well (172.6 mg per well). In the case of disc diffusion method, 5 µl of AgNPs solution was applied to each disc (8.63 µg per disc). In MIC assay, three-fold dilution series was performed from a stock solution (1.726 mg/mL) to obtain 57.46, 19.15, 6.38, 2.13, 0.71, 0.24, 0.08 and 0.03 µg/mL. The effect of AgNPs on the growth kinetics was made using growth curve and similar concentrations (6.38, 2.13, 0.71, 0.24, 0.08 and 0.03 µg/mL).

**Synergistic effect of AgNPs, Etoh extract**
For synergistic effect using well diffusion assay, a 50 µl of of each AgNPs and Etoh extract -DMSO solutions was combined to obtain a concentration of 86.28 µg and 5 mg, respectively. In disc diffusion synergistic assay, ready-made antibiotics discs were used as follow; amikacin.(30mcg), nitrofurantoin (300mcg) gentamicin (10mcg) trimethoprim/sulphamethoxazole (1.25/23.75mcg), chloramphenicol (30mcg) amoxicillin (20/10 mcg) and micafungin (1mg) caspofungin (5mg). To each of these antibiotics, in the case of dual synergistic effect 10 µl of either AgNPs or Etoh extract was used whereas in case of three-combined substances, each tested antibiotic and antifungal disc, 5 µl of undiluted Etoh extract (5µg per disc) and/or 5 µl of AgNPs solution (8.63µg/disc) were applied together (Klančnik *et al.,* 2010).

**Determination of minimum inhibitory concentration (MIC)**

Minimum inhibition concentration (MIC) values were determined using 96-well microtiter plates. Three-fold dilution was prepared. Suspensions of standard microorganisms adjusted to equivalent to 0.5 McFarland standards (approximately $10^8$ CFU/mL for bacteria), were inoculated onto the microplates. The growth of the microorganisms was monitored by sub cultured of each well content on nutrient agar. The MIC values were defined as the lowest concentrations of the plant Etoh, NPs, Etoh extract+NPs to inhibit the growth of microorganisms (Massa et al., 2018).

**RESULTS AND DISCUSSIONS**
**Biosynthesis and characterization of AgNPs:**
The fungal strain namely *Tritirachium oryzae* W5H; was previously isolated from this laboratory by Al-limoun et al. (2019). The colour of the culture filtrate with silver nitrate solution changed to intense brown after 72 h. of incubation, whereas, the control (silver nitrate salt free solution) did not show any colour change (Al-limoun et al., 2019; Qaralleh et al., 2019; Khleifat et al., 2019). The UV–vis spectrum showed a Surface Plasmon Resonance (SPR) peak of silver nanoparticles at 425 nm (Fig. 1). Morphological characterization of AgNP swas conducted using SEM which confirmed the presence of silver ions in the solution (Fig. 2). Different fungal species have been already applied to the synthesis of AgNPs such *as Aspergillus terreus, Pestalotiopsis sp., Pimelea columellifera subsp. Pallida, Aspergillus clavatus, Trichoderma harzianum, Penicillium aculeatum, Candida albicans, Fusarium verticillioides, and Emericella nidulans* (Li et al., 2011; Guilger et al., 2017; Rahimi et al., 2016; Mekkawy et al., 2017). This is the first report on the antimicrobial activity test on *A. graveolens ethanolic* extract that was synergized with of AgNPs.

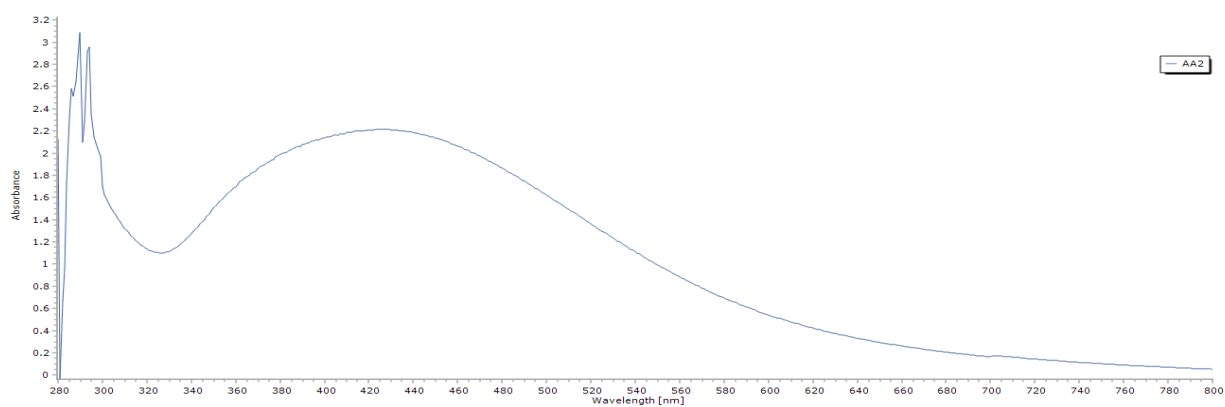

**Figure1. Ultraviolet-visible spectra of the solution of 1 mM AgNO3 after bioreduction by *Tritirachium oryzae* W5H at 33°C**

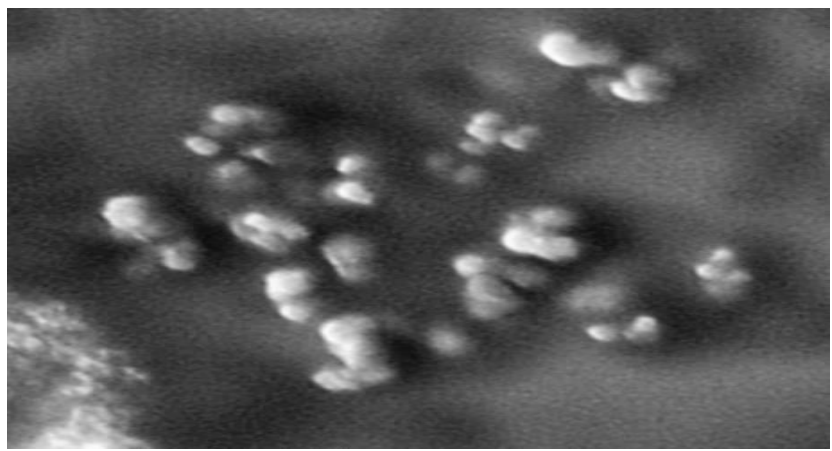

**Figure 2** Representative scanning electron microscopy (SEM) micrograph of Biosynthesized AgNPs by Fungal strain *Tritirachium oryzae* W5H.

**Antimicrobial activity of AgNPs, *A. graveolens* ethanolic extract and the synergistic effect between Etoh:AgNPs (1:1) using well diffusion assay.**

The antimicrobial activity of AgNPs and *A. graveolens* ethanolic extract was evaluated using well diffusion method against three gram negative strains (*E. aerogenes, Salmonella* sp. and *E. coli*)and one yeast strain (*C. albicans*). All strains tested except against *E. aerogenes* showed moderate antimicrobial activity toward AgNPs and *A. graveolens* ethanolic extract whereas *E. aerogenes* exhibited a 17 mm zone of inhibition (table 1). The antibacterial activity of *A. graveolens* ethanolic extract was higher than the antibacterial activity of AgNPs while the candidawas more susceptible to AgNPs (13 mm) and higher than that of *A. graveolens* ethanolic extract (7 mm). The results also showed that there were no remarkable increasing in antimicrobial activity when AgNPs was combined with *A. graveolens* ethanolic extract at equal proportion indicating an additive effect between them.

The antimicrobial activity was evaluated qualitatively using MIC. As shown in table 2, MICs of AgNPs against *E. aerogenes*, *Salmonella* sp., *E. coli* and *C. albicans* were 2.13, 19.15, 0.08 and 6.38 µg/mL, respectively, while the MICs of *A. graveolens* ethanolic extract against the same bacteria were 4, 366, >3300 and 40 µg/mL, respectively. The MIC values at concentration less than 19.15 and 40 µg/ml indicating the potent bacteriostatic effect of AgNPs and *A. graveolens* ethanolic extract. The synergistic effect of AgNPs: *A. graveolens* ethanolic extract was also evaluated against same pathogens. As shown in table 2, the MIC values were decreased when AgNPs was tested in combination with the *A. graveolens* ethanolic extract against *Salmonella* sp and *E.coli* (0.236:4.53 and 0.0789:1.5 µg/mL, respectively). *C. albicans* was more susceptible to AgNPs as compared to the tested bacteria. It showed 13 mm inhibition zonefollowed by *E. aerogenes* (12 mm), *E. coli* (11 mm) and *Salmonella* spwas the most resistant one (8 mm). Panáček et al. (2009) showed that AgNPs possess antifungal activity against *Candida* sp. at the concentration as low as 0.21 mg/L. Joanna et al., (2018) reported significant antifungal activity of AgNPs against *Alternaria alternate, Boeremia strassesi, Colletotrichum dematium, Colletotrichum fuscum, Cylindrocarpon destructans, Diaporthe eres, Diplocereus hypericinum, Fusarium equiseti, Fusarium oxysporum, Phylloticta plantagnus, Rhizoctonia solani, Sclerotinia sclerotiorum, Fusarium avenaceum, Fusarium avenaceum*. Preliminary study performed in our laboratory showed that AgNPs possesses broad-spectrum antibacterial activity (Tarawneh et al., 2011; Khleifat et al., 2019; Qaralleh et al., 2019). Guzman et al. (2012) showed that AgNPs synthesized chemically exhibited broad-spectrum antibacterial activity with maximum activity against *Escherichia coli, Pseudomonas aeruginosa* and *Staphylococcus aureus*.

**Table 1.** Antibacterial activity of *A. graveolens* ethanolic extract, AgNPs and *A. graveolens* ethanolic extract:AgNPs (1:1) on the tested bacterial strains using well diffusion method. Data are expressed as mean ± SD

| Bacterial strain | AgNPs | Etoh extract | AgNPs+Etoh extract |
|---|---|---|---|
| *E. aerogenes* | 12±0 | 17±0.5 | 18±0.5 |
| *Salmonella* | 8±0 | 13±0.5 | 14±0.5 |
| *E. coli* | 11±0 | 12±0 | 10±0 |
| *C. albicans* | 13±0 | 7±20 | 11±0 |

Each well contains 10 µg of *A. graveolens* ethanolic extract, 172.56 µg AgNPs or 5 µg Etoh: 86.28 µg AgNPs, AgNPs: silver nanoparticles.

Table 2. Minimum Inhibitory concentration (MIC) of AgNPs, *A. graveolens* ethanolic extract and combination of AgNPs: *A. graveolens* ethanolic extract

| Bacterial strain | NPs (µg/mL) | Etoh extract (µg/mL) | NPs+Etoh extract (µg/mL) |
|---|---|---|---|
| *E. aerogenes* | 2.13 | 4 | 57.52:1100 |
| *Salmonella* | 19.15 | 366 | 0.236:4.53 |
| *E. coil* | 0.08 | >3300 | 0.0789:1.5 |
| *C. albicans* | 6.38 | 40 | >172.56:>3300 |

The potential of using AgNPs in antimicrobial therapy is due to its multiple mode of bactericidal action. These mechanisms involve cell wall, cell membrane and key enzymes such as those engaged in ATP production, protein synthesis and DNA replication (Dorau et al., 2004; Duncan 2011; Duran et al., 2016; Tarawneh et al., 2009; Jung et al., 2008; Rauwel et al., 2015; Rahimi et al., 2016; Waghmare et al., 2015). Morones et al. (2005) described that the proton motive force of the membrane in *E. coli* dissipates when silver nanoparticles favorably bound and localized on the membrane of bacterial cells. In addition, the pitting of the cell membranes by silver nanoparticles causes a rise in permeability and lastly to the death of cell (Shahverdi et al., 2007; Lok et al., 2007). In particular, AgNPs+ attracted by the surface negative charge of the cell, adheres and penetrate the cell membrane. The adherent AgNPs inactivate the sulfhydryl groups in the cell wall, denature proteins and disrupt lipids on the cell membrane. Subsequently, this induce autolysis of the cell and spell out the cell contents. The intracellular AgNPs either bind to DNA and inhibit replication or disrupt electron transport chain and inhibit ATP production. Moreover, AgNPs have been found to facilitate the production of reactive oxygen species and induce oxidation damage.

In this study, *A. graveolens* ethanolic extract showed maximum inhibitory activity against *E. aerogenes* (17 mm) followed by *Salmonella* sp (13 mm) and *E. coli* (12 mm) while *C. albicans* was the most resistant strain (7 mm). Recent report showed that the ethyl acetate fraction of *A. graveolens* has moderate antibacterial activity against *Listeria monocytogenes* ATCC 19115 (MIC=0.312 mg/mL) (Ramdane et al., 2017).*A. graveolens* Etoh extract rich with 6-oxocyclonerolidol was reported with fungicidal activity against *Alternaria* sp. and *P. expansum* (Znini et al., 2011).

AgNPs synthesized by microorganisms have been approved to be good candidates for finding as antimicrobial agents. Recent researches showed that AgNPs mediated fungi synthesizes possessed broad spectrum antimicrobial activity against Gram-positive (*Bacillus subtilis, Staphylococcus aureus, Streptococcus faecalis*) and Gram negative (*Escherichia coli, Klebsiella pneumonia, Proteus mirabilis, Pseudomonas fluorescens, Pseudomonas aeruginosa,* and *Salmonella typhi*) bacteria (Barapatre et al., 2016; Verma et al., 2010; Rajendran et al., 2012; Ma et al., 2017). Pereira et al., 2014 showed that the AgNPs synthesized from *P. chrysogenum* exhibited stronger antibacterial activity comparing to fluconazole. AgNPs antimicrobial activity is correlated with several aspects including the particle size, shape and concentration (Burda et al., 2005; Bhainsa and D'Souza, 2006). As the size of the particles decrease, the antibacterial activity increased. This is due to the smaller size particles diffuse easily through the membrane comparing to the larger ones (Parthasarathy et al., 2015). It has been found that the particles shape (Kitching et al., 2015) also influences the antibacterial activity. Complete inhibition activity for the triangular AgNPs was at 1 mg whereas the spherical and rod AgNPs the inhibition activities were reported at 12.5 and 50-100 mg, respectively (Ramalingmam et al., 2015). The difference of antimicrobial activity between Gram positive and Gram-negative bacteria could be due to their variations in structure of cell wall. Gram-positive bacteria have a thick peptidoglycan layer, while the peptidoglycan layer in the Gram negative bacteria is thin but is surrounded by a lipid layer (Joanna et al., 2018).

**Synergistic effect of AgNPs and/or *A. graveolens* ethanolic extract combined with different antibiotics**

Antibiotic-resistant bacteria continue to challenge therapeutic practitioners. New model of antibiotics are continue expected due to the rise in the occurrence of infections caused by multidrug-resistant (MDR), methicillin-resistant, vancomycin-resistant, fluoroquinolone-resistant and carbapenem-resistant bacteria (Sanglard et al., 2009). Several strategies have been established to overcome the antibiotics resistance problem. One of the most common strategy is using active components from the natural sources. In addition, the use of NPs is considered as one of the most promising strategy in drug development. Combination therapy that might provide agent with multiple target sites are highly advantageous in reducing the antibiotics resistance problems.

The synergistic effect of AgNPs and/or *A. graveolens* ethanolic extract combined with 6 different conventional antibiotics was evaluated using disc diffusion method. The synergy effect against all tested pathogens was evaluated by an increase in fold area (IFA) of individual antibiotics and were calculated as $C=B^2-A^2/A^2$, where, A and B are the inhibition zones (mm) obtained for antibiotic alone and antibiotics plus AgNPs or antibiotics plus either ethanolic extract or combination of three (A +NPs +Ethanolic extract), respectively. In case of no zone of inhibition, diameter of the disk (6 mm) was considered for the calculation ( Abboud et al., 2009; Padalia et al., 2015).

As shown in table 3, the most effective antibiotic tested against *E. coli* was Gentamicin (21 mm) followed by amikacin (18 mm) in which both were not affected by the combination with AgNPs (no increase in IFA was reported). Moreover, IFAs of Gentamicin and amikacin combined with ethanolic extract and or AgNPs were 0.2 indicating weak synergistic effect between them. In addition, the most resistant strains to the tested antibiotics were Nitrofurantion and Amoxicillin. Among these, only Nitrofurantion showed synergistic effect when it was combined with AgNPs and with AgNPs: ethanolic extract since the increase in the IFA was by up to 1.0 and 0.7 respectively. This enhanced antibacterial activity reveals the noticeable solicitation in the field of nanomedicine. Chloramphenicol and Trimethoprim which exhibited moderate antibacterial activity against *E. coli* (13 and 12 mm, respectively) showed no synergistic effect except when Chloramphenicol was combined with AgNPs (IFA 0.5).

*E. aerogenes* was resistant to Amoxicillin while it was sensitive to amikacin, Gentamicin and Trimethoprim. As shown in table 4, no remarkable synergistic effect was reported when these antibiotics were combined with AgNPs or AgNPs: *A. graveolens* ethanolic extract. The exception of these are the increasing in IFA by 1 fold and 0.8 fold when amikacin was combined with Etoh extract and when Nitrofurantion was combined with AgNPs: *A. graveolens* ethanolic extract, respectively. IFA, designate an increase in fold area of individual antibiotics and were calculated as $C=B^2-A^2/A^2$, where, A and B are the inhibition zones (mm) obtained for antibiotic alone and antibiotics plus AgNPs or antibiotics plus Etoh of plants or combination of three (A +NPs +Etoh), respectively. All experiments were made in triplicates except the antibiotics disc was the average of twice and standard deviations were negligible. In case of no zone of inhibition, diameter of the disk (6 mm) was considered for the calculation

**Table 3.** Antibacterial activity (inhibition zone mm) of the antibiotics, combination of antibiotics:AgNPs (1:1), antibiotics: *A. graveolens* ethanolic extract (1:1), and antibiotics:AgNPs: *A. graveolens* ethanolic extract (1:1:1) against *E. coli*

| Antibiotic(A) | A | NPs+A | IFA | Etoh extract | IFA | A+NPs+Etoh extract | IFA |
|---|---|---|---|---|---|---|---|
| Nitrofurantion | 6 | 9 | 1.0 | 6 | 0 | 8 | 0.7 |
| Chloramphenicol | 13 | 16 | 0.5 | 10 | - | 10 | - |
| amikacin | 18 | 14 | - | 20 | 0.2 | 20 | 0.2 |
| Trimethoprim | 12 | 6 | - | 15 | 0.5 | 14 | 0.3 |
| Gentamicin | 21 | 10 | - | 16 | - | 18 | - |
| Amoxicillin | 6 | 6 | - | 6 | - | 6 | - |

IFA, designate an increase in fold area of individual antibiotics and were calculated as $C=B^2-A^2/A^2$, where, A and B are the inhibition zones (mm) obtained for antibiotic alone and antibiotics plus AgNPs or antibiotics plus ethanolic extract of plants or combination of three (A +NPs + Etoh extract), respectively. All experiments were made in triplicates except the antibiotics disc was the average of twice and standard deviations were negligible. In case of no zone of inhibition, diameter of the disk (6 mm) was considered for the calculation

**Table 4.** Antibacterial activity (inhibition zone mm) of the antibiotics, combination of antibiotics:AgNPs (1:1), antibiotics: *A. graveolens* ethanolic extract (1:1), and antibiotics:AgNPs: *A. graveolens* ethanolic extract (1:1:1) against *Enterobacter aerogenes*

| Antibiotic(A) | A | NPs+A | IFA | Etoh extract | IFA | A+NPs+Etoh extract | IFA |
|---|---|---|---|---|---|---|---|
| Nitrofurantion | 16 | 9 | - | 6 | - | 22 | 0.8 |
| Chloramphenicol | 15 | 8 | - | 10 | - | 6 | - |
| amikacin | 20 | 15 | - | 30 | 1.0 | 20 | 0 |
| Trimethoprim | 18 | 6 | - | 15 | - | 15 | - |
| Gentamicin | 20 | 19 | - | 16 | - | 16 | - |
| Amoxicillin | 6 | 6 | 0 | 6 | 0 | 6 | 0 |

IFA, designate an increase in fold area of individual antibiotics and were calculated as $C=B^2-A^2/A^2$, where, A and B are the inhibition zones (mm) obtained for antibiotic alone and antibiotics plus AgNPs or antibiotics plus Etoh of plants or combination of three (A +NPs +Etoh), respectively. All experiments were made in triplicates except the antibiotics disc was the average of twice and standard deviations were negligible. In case of no zone of inhibition, diameter of the disk (6 mm) was considered for the calculation

**Table 5.** Antibacterial activity (inhibition zone mm) of the antibiotics, combination of antibiotics:AgNPs (1:1), antibiotics: *A. graveolens* ethanolic extract (1:1), and antibiotics:AgNPs: *A. graveolens* ethanolic extract (1:1:1) against *Salmonella* sp.

| Antibiotic(A) | A | NPs+A | IFA | Etoh extract | IFA | A+NPs+Etoh extract | IFA |
|---|---|---|---|---|---|---|---|
| Nitrofurantion | 6 | 11 | 2 | 16 | 6 | 6 | 0 |
| Chloramphenicol | 18 | 6 | - | 22 | 0.4 | 22 | 0.4 |
| amikacin | 18 | 13 | - | 20 | 0.2 | 17 | - |
| Trimethoprim | 6 | 6 | - | 22 | 12 | 20 | 10 |
| Gentamicin | 15 | 6 | - | 20 | 0.7 | 18 | 0.4 |
| Amoxicillin | 18 | 9 | - | 23 | 0.6 | 20 | 0.2 |

IFA, designate an increase in fold area of individual antibiotics and were calculated as $C=B^2-A^2/A^2$, where, A and B are the inhibition zones (mm) obtained for antibiotic alone and antibiotics plus AgNPs or antibiotics plus Etoh of plants or combination of three (A +NPs +Etoh), respectively. All experiments were made in triplicates except the antibiotics disc was the average of twice and standard deviations were negligible. In case of no zone of inhibition, diameter of the disk (6 mm) was considered for the calculation

Table 6. Antifungal activity (inhibition zone mm) of the antifungals, combination of antifungals:AgNPs (1:1),antifungals:AgNPs: *A. graveolens* ethanolic extract (1:1:1) against *C. albicans*

| Antifungal | conc | A | + 10 μl AgNps | IFA | + 5 μAgNps + 5μ Etoh extract | IFA |
|---|---|---|---|---|---|---|
| Caspofungin | 14 ng | 9 | 7 | - | 6 | - |
|  | 27 ng | 12 | 14 | - | 6 | - |
| Micafungin | 1 ng | 12 | 6 | - | 21 | 0.75 |
|  | 5 ng | 19 | 17 | - | - | - |

*Salmonella* sp. was resistant to Nitrofurantion and Trimethoprim whereas it was sensitive to Chloramphenicol, amikacin and Amoxicillin (18 mm, each). Among the 6 antibiotics tested (table 5), increases in IFA by 2 fold was reported when Nitrofurantion and AgNPswere combined. In addition, all antibiotics tested showed IFAs when they were combined with Etoh extract achieving increase in IFA by 6 and 12 folds for Nitrofurantion and Trimethoprim, respectively. In addition, 10 folds increase in IFA was obtainedwith the combination of Trimethoprim:AgNPs: Etoh extract. It was reported that the synthesized silver nanoparticles enhanced the antibacterial property of carbicillin and moxifloxacin against clinically isolated pathogens in combined formulation (Althunibat et al., 2016; Gudikandula et al., 2017). Similar results were observed in synergistic effect of silver nanoparticles with commercial antibiotics (ß-lactams, glycopeptides, aminoglycosides and sulphonamides) (Thangapandiyan and Prema, 2012).

**Synergistic effect of AgNPs and/or *A. graveolens* ethanolic extract combined with Caspofungin and Micafungin against *C. albicans***

The synergistic effect of AgNPs with two antifungal agents was tested using disc diffusion method. Caspofungin and Micafungin were used in this study at different concentration. As shown in table, Micafungin showed stronger antifungal activity than Caspofungin. Maximum inhibition zone of 19 mm was reported for Micafungin at 5 ng while maximum inhibition zone of 12 mm was reported for Caspofungin at 27 ng. On the other hand, there were no synergistic effect between these two antifungal agents combined with either AgNPs or *A. graveolens* ethanolic extract. The exception of this is the combination of Micafungin with 5 μAgNps + 5μ Etoh extract (IFA equal to 0.75). The use of echinocandins (caspofungin and micafungin) is the drug of choice for the treatment of invasive candidiasis (Simon et al., 2013; Zeidan et al., 2013; Majali et al., 2015). These agents are ß-1,3-d-glucan synthase inhibitors. In which this enzyme essential for the synthesis of b-1,3-d-glucan polymers in the cell wall of several fungi (Rahimi et al., 2016) reported a potent synergistic effect of AgNPs with Caspofungin and Micafungin at 4 μg/mL. However, bacterial growth conditions can be instrumental in how these microbes respond to any antibacterial agents (Khleifat, 2007; Khleifat et al., 2007; Abboud et al., 2010; Khleifat et al., 2015).

**CONCLUSION**

The potent synergistic effect of *A. graveolens* ethanolic extract and/or AgNPs with the conventional antibiotics seems to be novel in inhibiting antibiotics resistant bacteria. Further pharmacological assays including in vivo study and cytotoxicity should be carried out to confirm the novelty of these combinations in treating UTIs. Combination therapy has been found to reduce the effective dose of antibiotics in the treatment of infections. This reduces the side effects and the toxicity of the antibiotic (Gibbons et al., 2003; Yap et al., 2014). Furthermore, combination therapy against resistant bacteria may have different mechanism of action and it may lead to find new agent that might help in microbial resistance management.